\documentclass{article}
\usepackage{spconf,amsmath,amssymb,graphicx,hyperref}

\title{Exploring Second-Order Pattern Recognition in Speaker Recognition}

\name{Yanze Xu$^{1}$, Wenwu Wang$^{1}$, Mark D. Plumbley$^{2}$}
\address{$^{1}$Centre for Vision, Speech and Signal Processing, University of Surrey, Guildford, UK\\
$^{2}$Department of Informatics, King's College London, London, UK}

\begin{document}

\maketitle

\begin{abstract}

In traditional pattern recognition tasks, neural networks are trained to recognise human-defined patterns (e.g.\ audio categories) in model inputs (e.g.\ audio). Meanwhile, some Explainable AI (XAI) methods explain latent patterns characterising the network's recognition of inputs as human-defined patterns; this work calls these latent patterns second-order patterns and proposes to discover them. Accordingly, we apply a hierarchical clustering algorithm to analyse whether our speaker recognition network's representations learned from known utterances naturally form hierarchical clusters. Each resulting cluster is a second-order pattern that characterises a context in our network's recognition of the known utterances as speaker identities. All discovered second-order patterns are then interpreted using the Hierarchical Cluster-Class Matching (HCCM) method.

Moreover, we propose a new task, second-order pattern recognition, to identify which of the discovered second-order patterns characterising the recognition of known utterances also apply to unseen utterances, thereby characterising the recognition of unseen utterances. Accordingly, we design the Hierarchical Cluster Navigation and Assignment (HCNA) method. HCNA recognises a second-order pattern as applying to an unseen utterance when the utterance's network representation lies within the extrapolation space of the cluster regarded as that second-order pattern. Experimental results demonstrate that the extrapolation space introduced in HCNA substantially improves task performance.

\end{abstract}


\section{Introduction}
\label{sec:intro}

Humans define patterns in audio or images such as speaker identities, phonemes, and facial identities. Traditional pattern recognition tasks aim to automatically recognise the human-defined patterns in data~\cite{jain2000statistical}. Most modern pattern recognition models are implemented as neural networks trained to learn a non-linear mapping from observed data directly to human-defined patterns~\cite{krizhevsky2012imagenet, chung2018voxceleb2}. However, neural networks operate largely opaquely, posing risks when deployed in real-world applications. The emerging field of Explainable Artificial Intelligence (XAI)~\cite{xu2026explainable2}
aims to explain the decision-making processes of trained models, especially neural networks. Some XAI techniques~\cite{zhou2016learning, selvaraju2017grad} can explain latent patterns that characterise the network's recognition of inputs as human-defined patterns. Yet, these latent patterns discovered by XAI techniques are rarely intended to be recognised from model inputs in traditional pattern recognition tasks.

Conceptually, studies in cognitive psychology consider a person's belief about an object a \textit{first-order belief}, and a person's belief about another person's belief a \textit{second-order belief}~\cite{perner1985john}. Analogously, we propose to regard a network's ``belief'' that inputs can be recognised as certain human-defined patterns as a first-order belief, and an XAI technique's ``belief'' that some latent patterns characterise a network's ``belief'' (i.e.\ recognition of inputs) as a second-order belief. Under this anthropomorphic analogy, we also call the human-defined patterns in the network's first-order belief \textit{first-order patterns}, and the latent patterns in the XAI technique's second-order belief \textit{second-order patterns}.




Two research questions are then asked. First, does the real world contain a number of second-order patterns that remain to be discovered using XAI techniques? Secondly, once second-order patterns characterising the network's recognition of some known data have been discovered, can we automatically recognise which of these patterns also apply to unseen data, i.e., also characterise the same network's recognition of the unseen data? We call this new pattern recognition task \textit{second-order pattern recognition}.


This work aims to realise the second-order pattern recognition task based on the speaker recognition network~\cite{nagrani2017voxceleb, chung2018voxceleb2}. Specifically, we first prepare a collection of utterances and extract their representations (i.e.\ speaker embeddings) from a trained speaker recognition network. We next apply a hierarchical clustering algorithm~\cite{mullner2011modern}, Single-Linkage Clustering (SLINK)~\cite{gower1969minimum, sibson1973slink}, previously applied in the XAI context~\cite{xu2026explainable1}, to analyse whether these speaker embeddings naturally form clusters with hierarchical relationships (i.e.\ hierarchical representation clusters). Each resulting cluster is a second-order pattern that, in SLINK's ``belief'', characterises a context in our network's recognition of the prepared utterances, in which our network discriminates subsets of these utterances.


The above second-order patterns discovered by SLINK can be made understandable to humans through semantic interpretation. Specifically, we apply the existing \emph{Hierarchical Cluster-Class Matching} (HCCM) method~\cite{xu2026explainable1} to match SLINK's hierarchical representation clusters with predefined semantic classes in a one-to-one manner. Each cluster regarded as a second-order pattern is interpreted according to the semantic class with which it is matched.


Although the above second-order patterns only characterise our network's recognition of the prepared utterances, hereafter referred to as known utterances, these patterns are intended to be recognised from unseen utterances in the proposed second-order pattern recognition task. Accordingly, we introduce another collection of unseen utterances that have not been analysed by SLINK. We then design the \emph{Hierarchical Cluster Navigation and Assignment} (HCNA) method, which identifies a second-order pattern associated with known utterances as applicable to an unseen utterance by determining whether the unseen utterance's representation belongs to the cluster regarded as that second-order pattern.

Following McInnes et al.~\cite{mcinnes2017hdbscan}, an unseen representation belongs to a cluster when it lies within the \textit{empirical space} defined by the representations in that cluster. Instead of using the empirical space, HCNA estimates the \textit{extrapolation space}~\cite{haley1992extrapolation} for each cluster, defined as the space to which the representations in a cluster can generalise. In more detail, HCNA estimates this extrapolation space by introducing the \emph{Gradient Cost}, which quantifies the gradient expenditure required to optimise a unit difference in the loss function. This Gradient Cost is then used to adjust the cluster's empirical space into its extrapolation space, within which the membership of an unseen utterance's representation is determined.





Overall, the second-order pattern recognition task is operationalised primarily through the HCNA method, while its realisation also relies on SLINK and HCCM to discover and interpret second-order patterns. These three methods form a complete second-order pattern recognition system. Several evaluation metrics are introduced to assess this system. Experimental results show that using the extrapolation spaces estimated by HCNA substantially improves system performance on the second-order pattern recognition task compared with using the empirical spaces alone.





\section{Related Work}
\label{sec:related}

Most speaker embeddings are not understandable to humans. To address this, XAI studies infer the semantic information encoded in speaker embeddings by training downstream classifiers to map speaker embeddings to semantic classes related to gender, nationality, or phonetic information~\cite{wu2024explainable, luu2020leveraging, luu2022investigating}. Successful downstream classification implicitly indicates that the semantic information is present in the embeddings. In comparison, our work directly and explicitly decodes the semantic information encoded in speaker embeddings by first analysing second-order patterns in the embeddings (i.e.\ representation clusters) using SLINK and then associating these patterns with semantic classes using HCCM.


Similar works apply clustering algorithms to network representations of known data, treating the resulting clusters as pseudo-labels rather than second-order patterns, and then predicting whether unseen data belong to these pseudo-labels~\cite{caron2018deepcluster, dopierre2020fewshot, vangansbeke2020scan}. In comparison, second-order patterns in this work are conceptualised with reference to the second-order belief of an XAI technique, with the practical role of characterising the network's recognition, whereas the pseudo-labels in these related works are neither conceptualised in a similar way nor assigned a similar role. Moreover, second-order patterns are interpreted using HCCM, whereas the pseudo-labels in related works are not further interpreted.


\section{METHODOLOGY}
\label{sec:method}
\subsection{Discovering and Interpreting Second-Order Patterns}

Let $f(\cdot)$ denote a trained speaker recognition network, which maps an utterance $x$ to a speaker embedding $a=f(x)$. An instantiation of the network's first-order belief introduced in Section~\ref{sec:intro} is that, in network $f$'s ``belief'', $x$ can be recognised in terms of speaker identities using the learned representation $a$. We prepare a set of known utterances $X=\{x_i\}$, which is independent of the data used to train $f$. The network representations of $X$ are denoted by $A=\{a_i\}$, where $a_i=f(x_i)$.

We then apply the hierarchical clustering algorithm~\cite{mullner2011modern}, SLINK~\cite{gower1969minimum,sibson1973slink}, to representations $A$. SLINK analyses a hierarchical representation organisation in which representations $A$ naturally form hierarchical clusters, denoted by $\mathcal{H}=\{h_1,\ldots, h_i \ldots\}$, where $h_i$ is the $i$-th hierarchical representation cluster containing the indices of all representations belonging to it. Each cluster $h_i$ has a sibling cluster in $\mathcal{H}$, such that representations in $h_i$ are discriminated from representations in its sibling cluster. This provides an instantiation of an XAI technique's second-order belief. In SLINK's ``belief'' about the network $f$'s ``belief'', $h_i$ is a latent pattern (i.e.\ a second-order pattern) characterising a certain context in the network $f$'s recognition of known utterances $X$, in which the network locally organises and discriminates two subgroups of utterances' representations (i.e.\ $h_i$ and its sibling cluster).

To offer semantic interpretations of all clusters in $\mathcal{H}$ regarded as second-order patterns, the Hierarchical Cluster-Class Matching (HCCM) method~\cite{xu2026explainable1} is used to determine which predefined semantic classes (i.e.\ classes pre-labelled for utterances $X$ and their representations $A$) match the clusters in $\mathcal{H}$ under a one-to-one correspondence. Each cluster is then interpreted by the predefined semantic class that achieves the highest matching degree with that cluster. As more predefined semantic classes become available, more $\mathcal{H}$'s clusters regarded as second-order patterns can be semantically interpreted. The matching degree between a cluster $h \in \mathcal{H}$ and a semantic class $c$, where $c$ is an index set containing all representations whose speaker identities belong to that semantic class, is quantified below using the L-score~\cite{xu2026explainable1}:
\begin{equation}
    L(h,c)
    =
    \frac{|h\cap c|}
    {\max(|h|,|c|)}.
\end{equation}
At this stage, SLINK discovers second-order patterns that characterise the network's recognition of known utterances, while HCCM provides interpretations for these patterns. In the next section, we propose the \emph{Hierarchical Cluster Navigation and Assignment} (HCNA) method to determine whether these second-order patterns also apply to an unseen input, i.e., characterise the network's recognition of that unseen input.



\subsection{HCNA Part I - Candidate Path}
\label{sec:path}
The HCNA method first identifies candidate second-order patterns in $\mathcal{H}$ that may apply to an unseen utterance. For an unseen utterance $x_{\mathrm{unseen}}$, which is independent of known utterances $X$ and data used to train $f$, the unseen representation is obtained as $a_{\mathrm{unseen}}=f(x_{\mathrm{unseen}})$. HCNA finds the index of the nearest known representation in $A$ as follows:
\begin{equation}
    i^*
    =
    \arg\min_i \mathrm{dist}(a_{\mathrm{unseen}},a_i),
    \label{eq:nn}
\end{equation}
\noindent where $\mathrm{dist}(\cdot,\cdot)$ denotes the Euclidean distance metric. The corresponding representation $a_{i^*}$ is therefore the known representation in $A$ nearest to $a_{\mathrm{unseen}}$. Within the hierarchical representation organisation $\mathcal{H}$, all clusters containing $a_{i^*}$ are nested and form a path $h^{(1)}\rightarrow h^{(2)}\rightarrow\cdots\rightarrow h^{(m)}$, where $m$ is the number of clusters containing $a_{i^*}$. We treat all clusters on this path as candidate second-order patterns. HCNA subsequently navigates through these candidates one by one to recognise those that actually apply to the utterance $x_{\mathrm{unseen}}$.

Recognising a candidate second-order pattern $h^{(t)}$ as applicable to $x_{\mathrm{unseen}}$ is equivalent to determining whether $a_{\mathrm{unseen}}$ belongs to the corresponding cluster. To determine this cluster membership, the \textit{birth distance} of $h^{(t)}$, denoted by $d_{\mathrm{birth}}^{(t)}$, is defined as the minimum distance between $h^{(t)}$ and its sibling cluster, i.e., the distance between their closest representations. Based on this definition, the \textit{empirical space} of $h^{(t)}$ is the space covered by its representations within the range defined by $d_{\mathrm{birth}}^{(t)}$. Following McInnes et al.'s method~\cite{mcinnes2017hdbscan}, $a_{\mathrm{unseen}}$ is recognised as belonging to $h^{(t)}$ if and only if $a_{\mathrm{unseen}}$ lies within $h^{(t)}$'s empirical space, equivalently when $\mathrm{dist}(a_{i^*},a_{\mathrm{unseen}})<d_{\mathrm{birth}}^{(t)}$.

\subsection{HCNA Part II - Gradient Cost and Extrapolation}
\label{sec:gc}

HCNA determines cluster membership for an unseen utterance's representation using its own criterion, thereby identifying the applicable candidate second-order patterns. In particular, HCNA examines whether the unseen representation lies within the extrapolation spaces of the clusters, rather than within their empirical spaces.

We now compute the extrapolation space of the cluster $h^{(t)}$ regarded as a candidate second-order pattern step by step. Specifically, we first introduce a new quantity termed \emph{Gradient Cost}, defined as the amount of gradient adjustment required per unit difference in the loss function. In our experimental setting, the examined speaker recognition model $f$ is trained using the angular prototypical loss function~\cite{chung2020defence}, which minimises angular differences. Hence, this work formulate Gradient Cost as the amount of gradient adjustment required for optimisation per unit angular difference:
\begin{equation}
    GC_{\mathrm{unseen}}
    =
    \left\|
    \frac{\|a_{\mathrm{unseen}}\|_2}{\theta_{\mathrm{unseen}}}
    \frac{\partial L_{\mathrm{unseen}}}{\partial a_{\mathrm{unseen}}}
    \right\|_1,
    \label{eq:gcunseen}
\end{equation}

\noindent where $L_{\mathrm{unseen}}=\frac{1}{2}\theta_{\mathrm{unseen}}^2$ is the optimisation loss function for reducing the angular difference between $a_{\mathrm{unseen}}$ and its nearest known representation $a_{i^*}$, and $\theta_{\mathrm{unseen}}$ denotes this angular difference, given by $\arccos\left(\frac{a_{\mathrm{unseen}}^\top a_{i^*}}{\|a_{\mathrm{unseen}}\|_2\|a_{i^*}\|_2}\right)$. The gradient $\frac{\partial L_{\mathrm{unseen}}}{\partial a_{\mathrm{unseen}}}$ indicates how $a_{\mathrm{unseen}}$ should be adjusted to reduce its angular difference from $a_{i^*}$, thereby moving it towards integration into $h^{(t)}$, which contains $a_{i^*}$. Dividing this gradient by $\theta_{\mathrm{unseen}}$ measures the gradient adjustment per unit angular difference, while multiplying by $\|a_{\mathrm{unseen}}\|_2$ compensates for the scale of the representation. Next, we quantify the \emph{Gradient-Cost Ratio} using $GC_{\mathrm{unseen}}$ as follows:
\begin{equation}
    GCR^{(t)}
    =
    \frac{GC_{\mathrm{unseen}}}
    {GC_{\mathrm{birth}}^{(t)}},
    \label{eq:gcr}
\end{equation}

\noindent where $GC_{\mathrm{birth}}^{(t)}$ is the Gradient Cost of integrating cluster $h^{(t)}$ with its sibling cluster, computed by minimising the angular difference between the closest pair of representations from these two clusters, which defines their birth distance. When $GCR^{(t)}>1$, meaning that integrating $a_{\mathrm{unseen}}$ into $h^{(t)}$ requires a greater Gradient Cost than integrating $h^{(t)}$ with its sibling cluster, we hypothesise that this indicates that $a_{\mathrm{unseen}}$ is relatively difficult to integrate into $h^{(t)}$. Conversely, when $GCR^{(t)}<1$, meaning that integrating $a_{\mathrm{unseen}}$ into $h^{(t)}$ requires a lower Gradient Cost than integrating $h^{(t)}$ with its sibling cluster, we hypothesise that $a_{\mathrm{unseen}}$ is relatively easy to integrate into $h^{(t)}$. Taken together, $GCR^{(t)}$ measures the relative difficulty of integrating $a_{\mathrm{unseen}}$ into $h^{(t)}$ compared with integrating $h^{(t)}$ with its sibling cluster.


Lastly, HCNA adjusts the birth distance of $h^{(t)}$ (i.e.\ $d_{\mathrm{birth}}^{(t)}$) according to $GCR^{(t)}$ to obtain a new distance $d_{\mathrm{extra}}^{(t)}$:
\begin{equation}
d_{\mathrm{extra}}^{(t)}
=
\frac{d_{\mathrm{birth}}^{(t)}}{GCR^{(t)}},
\label{eq:dextra}
\end{equation}
\noindent where a higher integration difficulty between $a_{\mathrm{unseen}}$ and $h^{(t)}$ contracts the empirical space of $h^{(t)}$ in the direction of $a_{\mathrm{unseen}}$, reducing its range from $d_{\mathrm{birth}}^{(t)}$ to $d_{\mathrm{extra}}^{(t)}$. Conversely, a lower integration difficulty expands the empirical space in that direction. Eventually, for the given $a_{\mathrm{unseen}}$, we define its own extrapolation space of $h^{(t)}$ as the space covered by all known representations in $h^{(t)}$ within distance $d_{\mathrm{extra}}^{(t)}$, such that the empirical space of $h^{(t)}$ adjusted using $GCR^{(t)}$ becomes the extrapolation space of $h^{(t)}$. Notably, $h^{(t)}$'s extrapolation space is not fixed, but instead varies with the presented unseen representation in our formulation.

Finally, the unseen representation $a_{\mathrm{unseen}}$ is regarded as lying within its own $h^{(t)}$'s extrapolation space once $d(a_{\mathrm{unseen}},a_{i^*})<d_{\mathrm{extra}}^{(t)}$. If this condition is satisfied, the candidate second-order pattern represented by cluster $h^{(t)}$, together with $h^{(t)}$'s interpretation provided by HCCM, is recognised as applying to the unseen utterance $x_{\mathrm{unseen}}$, thereby characterising model $f$'s recognition of $x_{\mathrm{unseen}}$. In this way, HCNA navigates through all candidate second-order patterns for $x_{\mathrm{unseen}}$ and progressively identifies the applicable ones.

\section{Experiments}
\label{sec:experiments}

\subsection{Experimental Setup}

We use the ResNetSE34L speaker recognition model published by Chung et al.~\cite{chung2020defence}, trained on the VoxCeleb2 development set~\cite{chung2018voxceleb2} using the angular prototypical loss~\cite{chung2020defence}. We use the VoxCeleb1 test set~\cite{nagrani2017voxceleb} as known utterances for discovering second-order patterns, and the VoxCeleb2 test set~\cite{chung2018voxceleb2} as unseen utterances. SLINK-analysed clusters containing fewer than 2000 representations are excluded to limit the number of second-order patterns. Predefined semantic classes related to gender and nationality are collected from the Internet for all examined utterances, with those of known utterances used by HCCM to interpret the discovered second-order patterns.

\subsection{Results}

Table~\ref{tab:results} compares the performance of the system (i.e.\ SLINK + HCCM + HCNA) on the second-order pattern recognition task under three HCNA settings. \emph{HCNA w/o extrap.} determines cluster membership based on the empirical space. \emph{HCNA GCR extrap.} relies on the extrapolation space estimated using the GCR defined in Section~\ref{sec:gc}. \emph{HCNA fixed extrap.} determines cluster membership by extrapolating the empirical space using a fixed factor instead of the GCR. 

The first evaluation metric in Table~\ref{tab:results}, \emph{extrapolation magnitude}, measures the average extent to which all clusters corresponding to applicable second-order patterns across unseen utterances are extrapolated beyond their empirical spaces. \emph{Path completeness} measures the average, across all unseen utterances, of the proportion of candidate second-order patterns (i.e.\ clusters) that are successfully recognised (i.e.\ traversed). Moreover, our HCCM method interprets some second-order patterns associated with known utterances using individual classes (e.g.\ \emph{UK}). Based on these interpretations, the third metric, \emph{Atomic semantic recall}, measures the average proportion of individual classes pre-labelled for unseen utterances that are also present in the HCCM interpretations of their applicable second-order patterns. \emph{Gender atomic recall} and \emph{nation atomic recall} measure the same metric using only individual classes related to gender and nationality, respectively.


\begin{table}[t]
\caption{Evaluation of the second-order pattern recognition system under different HCNA extrapolation settings.}
\label{tab:results}
\centering
\small
\setlength{\tabcolsep}{3.5pt}
\begin{tabular}{lccc}
\hline
Metric &
\begin{tabular}{c}
HCNA\\
w/o extrap.
\end{tabular}
&
\begin{tabular}{c}
HCNA\\
fixed extrap.
\end{tabular}
&
\begin{tabular}{c}
HCNA\\
GCR extrap.
\end{tabular}
\\
\hline
Extrap. magnitude   & --      & 7.16\% & 7.16\% \\
Path completeness         & 11.68\% & 24.05\% & \textbf{34.19\%} \\
Atomic recall    & 0.5119  & 0.5423  & \textbf{0.5669} \\
Gender atomic recall      & \textbf{98.73\%} & 98.58\% & 98.18\% \\
Nation atomic recall & 3.64\%  & 9.88\%  & \textbf{15.20\%} \\
\hline
\end{tabular}
\end{table}

In Table~\ref{tab:results}, path completeness and atomic semantic recall increase from 11.68\% and 0.5119 without an extrapolation mechanism, to 24.05\% and 0.5423 with a fixed extrapolation magnitude of 7.16\%, and then to 34.19\% and 0.5669 with GCR-based extrapolation of the same 7.16\% magnitude. Gender atomic recall remains approximately 98\%, whereas nation atomic recall increases from 3.64\% to 9.88\% and 15.20\%, with fixed and GCR-based extrapolation yielding the latter two values, respectively. These improvements demonstrate the benefits of adaptively adjusting the extrapolation spaces of clusters using the GCR, rather than uniformly enlarging their empirical spaces or leaving them unchanged.



\section{Conclusion}

This work realises a new task, second-order pattern recognition, built upon the speaker recognition network. In this task, second-order patterns that, according to an XAI technique's ``belief'', characterise our network's recognition of known utterances in terms of speaker identities are used to determine which of these patterns apply to unseen utterances. A baseline system integrating SLINK, HCCM, and HCNA is developed, with SLINK and HCCM used to discover and interpret second-order patterns from known utterances, and HCNA used to recognise their applicability to unseen utterances. HCNA introduces Gradient Cost and Gradient Cost Ratio (GCR) to estimate the extrapolation spaces of SLINK's hierarchical representation clusters, and examines whether an unseen utterance's representation lies within these spaces to identify the utterance's applicable second-order patterns. Experimental results show that HCNA's GCR-based extrapolation improves performance on the second-order pattern recognition task compared with no extrapolation, increasing path completeness from 11.68\% to 34.19\% and atomic semantic recall from 0.5119 to 0.5669. GCR-based extrapolation also outperforms fixed extrapolation under the same extrapolation magnitude of 7.16\%, demonstrating the benefit of adaptively extrapolating the empirical spaces of clusters using the GCR.

\section{Acknowledgement}
Mark D. Plumbley was supported by the Engineering and Physical Sciences Research Council (EPSRC) [grant number EP/Y028805/1]. For the purpose of open access, the authors have applied a Creative Commons Attribution (CC BY) licence to any Author Accepted Manuscript version arising.

\bibliographystyle{IEEEbib}
\bibliography{refs}

\end{document}